\documentstyle[psfig]{l-aa}

\def\simlt {\lower.5ex\hbox{$\; \buildrel < \over \sim \;$}}
\def\simgt{\lower.5ex\hbox{$\; \buildrel > \over \sim \;$}}

\def\ftools {{\em ftools}}
  
\def\net {$n_{\rm e}t$\ } 
\def\NH {$N_{\rm H}$\ } 
\def\HAlpha {H$\alpha$\ } 
\def\EM {$n_{\rm e} n_{\rm H} V/d^2$}
\def\Te {$T_{\rm e}$}

\def\densm {m$^{-3}$\ }

\begin{document}
   \thesaurus{ 2.01.4 ; 08.19.5; 09.01.1; 09.19.2; 13.25.4}

    \title{X-ray spectroscopy of the supernova remnant RCW 86} 
	\subtitle{A new challenge for modeling the emission from supernova remnants}

    \author{Jacco Vink \and Jelle S.  Kaastra \and Johan A. M. Bleeker}

    \offprints{J.Vink@sron.ruu.nl}

    \institute{Space Research Organization Netherlands, Sorbonnelaan 2 NL-3584
              CA Utrecht, The Nether\-lands }

\date{}

\maketitle

\begin{abstract}
We present an analysis of ASCA X-ray data of SNR RCW\,86. 
There appears to be a remarkable spectral variation over the remnant, 
indicating temperatures varying from 0.8~keV to $> 3$~keV. 
We have fitted these spectra with non-equilibrium 
ionization models and found that all regions are best fitted by
emission from a hot plasma underabundant in metals ($<0.25$ solar),
but in some cases fluorescent emission indicates overabundances of Ar and Fe. 
The ionization stage of the metals appears to be far from 
equilibrium, at some spots as low as 
$\log(n_{\rm e}t)$ (m$^{-3}$s) $\sim 15.3$. We discuss the physical reality
of the abundances and suggest an electron distribution with a supra-thermal 
tail to alleviate the strong depletion factors observed.
We argue that RCW\,86 is the result of a cavity explosion.

\end{abstract}

\section{Introduction}

RCW\,86 (also G315.4-2.3 and MSH 14-63) was for some time regarded as the 
historical remnant of the supernova (SN) of AD 185 (Clark \& Stephenson 1977). 
Recently, however, Chin \& Huang (1994) presented convincing evidence that 
the event recorded by the Chinese was not a SN at all.
This makes RCW\,86 just one of many other galactic supernova remnants (SNRs). 
Nevertheless, the ASCA X-ray spectra show that RCW\, 86 has 
some very special properties.

The radius of the remnant is $\sim$22\arcmin. Distance estimates were usually 
based on the notion that RCW\,86 was SN185. The most recent (kinematic)
distance estimate is 2.8~kpc (Rosado et al. 1996). 

\begin{figure*}[td]
%	\picplace{15cm}
	\psfig{figure=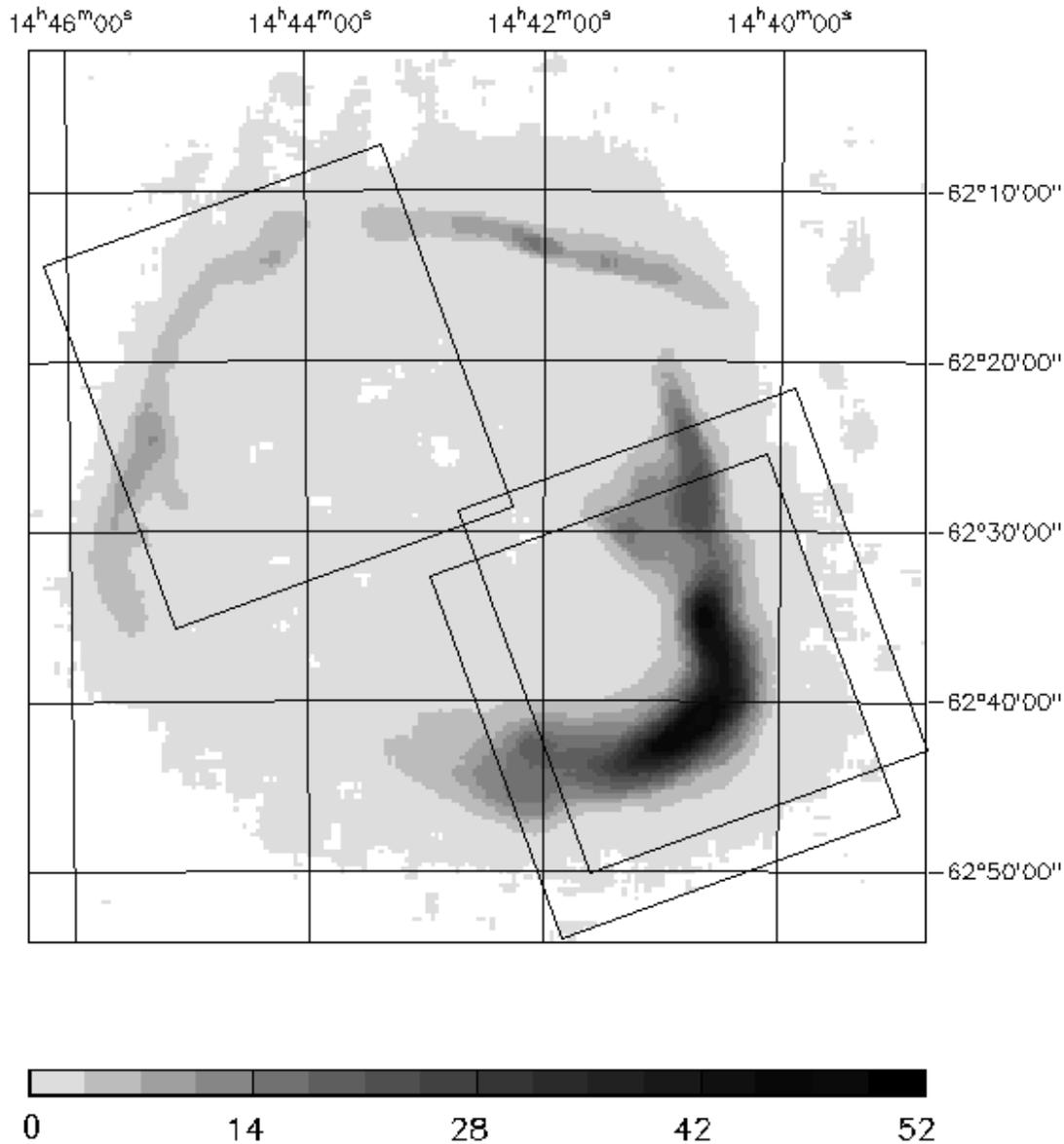,width=15cm}
  	\caption[]{
The ROSAT PSPC image of RCW\, 86 in the energy 
range 0.1-1~keV. The image is composed of three observations,
with a total integration time of 9ks.
The image was smoothed with a median filter of $7 \times 7$ pixels, 
the pixel size being 15\arcsec.
The grey scale is such that one can see a clear distinction between the faint
emission (the lightest grey) and the bright shell already known from the 
Einstein (Pisarski et al. 1984) and EXOSAT missions (Claas et al. 1989). 
Note that the satellite pointing was such that the Southwestern
region was observed with less sensitivity than the Northeastern region.
The three squares indicate the areas covered by the ASCA SIS instruments.
\label{image}
	}
\end{figure*}

\begin{figure*}[td]
	\psfig{figure=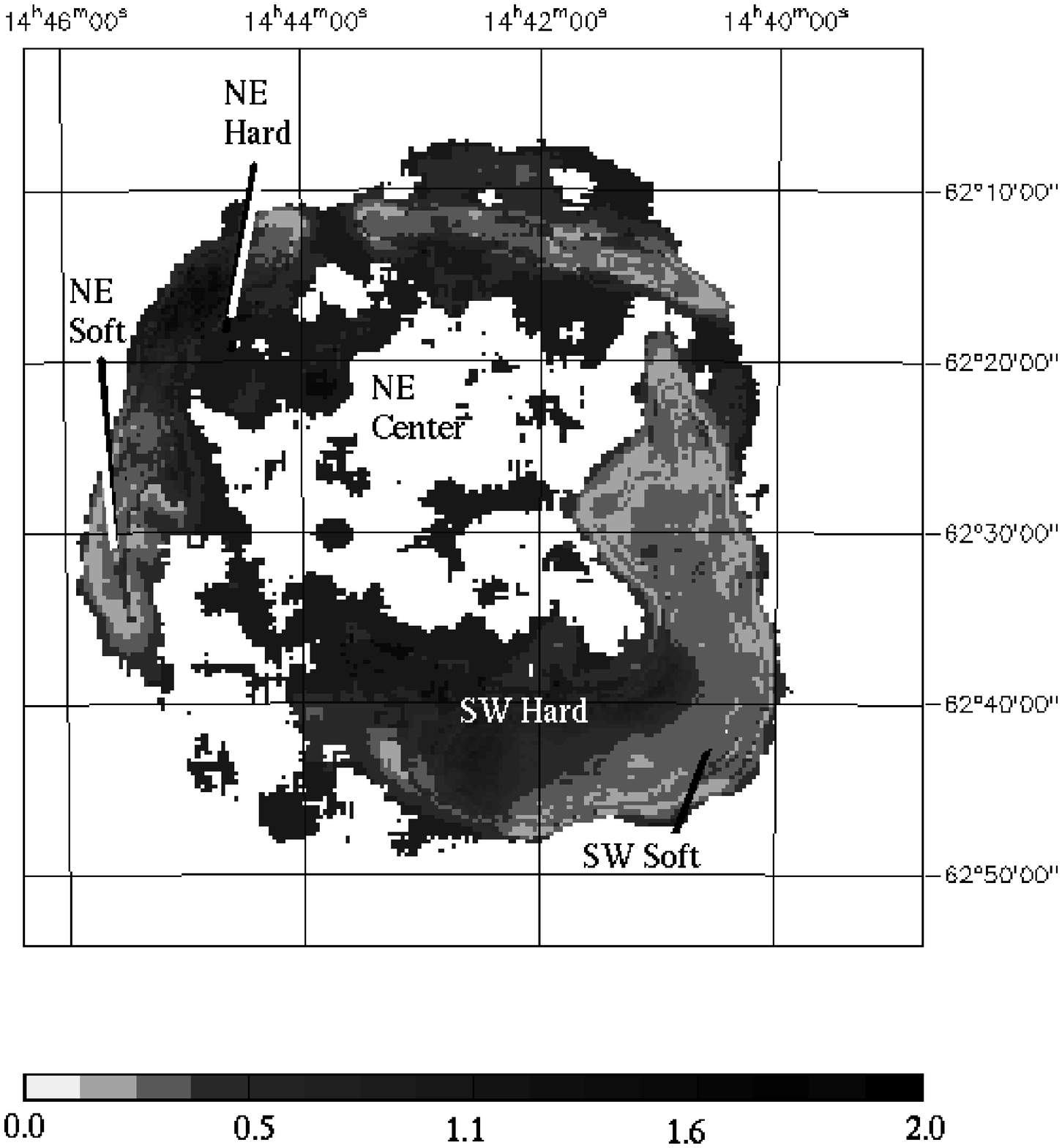,width=15cm}
  	\caption[]{
The PSPC ratio map, based on the same observations as Fig.~\ref{image}.
The images are composed of three observations,
with a total integration time of 9ks, 
The images were extracted in the 
energy ranges 0.1--1~keV and 1--2.1~keV and smoothed with a 
median filter of $7 \times 7$ pixels, the pixel size being 15\arcsec.
A higher ratio indicates a harder spectrum. The color scale is such that they 
represent our choice for selection of spectrum well.
\label{ratio}
	}
\end{figure*}

\section{The data}
Our analysis is mainly based on archival ASCA data (Tanaka et al. 1994). 
RCW\,86 was observed by ASCA in the PV phase on August 17, 1993. 
Three pointings were used. Two of the
pointings (together $\sim 14$~ks) cover the bright boomerang shaped area in 
the Southwest of the remnant.
The third observation ($\sim 14$~ks) covers the
Northeastern part of the remnant. The Northwestern part of the 
remnant is not covered. The area covered by ASCA is indicated in the ROSAT 
PSPC image Fig.~\ref{image}.

For our analysis we used spectra from all four instruments on board ASCA
(i.e. SIS 0\&1, GIS 2\&3). The spectral resolution of the SIS
instruments is better than that of the GIS (2\% vs. 8\% FWHM at 6~keV),
but the latter have the advantage of better sensitivity at high energies. 
SIS observations were made in 4 CCD faint and bright modes.

We reduced the data with the NASA/HEASARC \ftools\ package v3.6
using standard screening criteria. Background spectra were extracted from
the standard background data available at the HEASARC archive. 
The response matrices for the SIS spectra were generated with {\em sisrmg},
adding matrices of the various chips with the appropriate statistical weights
and auxiliary response files.

The angular resolution of the ASCA instruments ensures that potential 
contamination 
by the galactic ridge is negligible (Claas et al. 1989, Kaastra et al. 1992).

We also analyzed archival ROSAT PSPC data of RCW\,86.
Three pointings focus on the North and one on the
South of the remnant. The PSPC has the advantage of having a better spatial 
resolution (FWHM $\sim$ 30\arcsec), than ASCA (FWHM $\sim$ 60\arcsec), 
but the spectral resolution is poor. PSPC data were reduced with the 
\ftools\ package as well.

\section{Spectral analysis and interpretation}

\begin{table*}[td]
	\caption[]{
Plasma parameters for one and two temperature fits to the ASCA
spectra. The abundances are number abundances in fractions of solar 
abundances (Anders \& Grevese  1989). Numbers in brackets give the 90\% 
confidence range ($\Delta\chi^2 = 2.7$).\label{parameters}
	}
	{\scriptsize
  		\begin{flushleft}
     			\begin{tabular}{llllllllllll}
       				\hline\noalign{\smallskip}
   &\multicolumn{2}{l}{SW Soft}&\multicolumn{2}{l}{SW Hard}&\multicolumn{2}{l}{NE Soft}
   & \multicolumn{2}{l}{NE Center}&\multicolumn{2}{l}{NE Hard}\\
        			\noalign{\smallskip}
        			\hline\noalign{\smallskip}
%                                  SW Soft                 SW Hard               NE Soft           NE Center            NE Hard
\EM\,($10^{62}$m$^{-3}$kpc$^{-2}$)& 11   &             & 11    &            & 3    &            & 1.3  &            & 2.4   &       \\
$kT_{\rm e}$ (keV)               & 0.79 & (0.73--0.85) & 1.2   & (1.0--1.4) & 0.6  & (0.5--0.9) & 0.8  & (0.6--1.1) & 2.0   & (1.8--2.1)\\
\net ($10^{15}$m$^{-3}$s)        & 9.8  & (9.0-11)     & 5.6   &(5.0--6.1)  & 11   & (7--20)    & 2.6  & ($ < 8$)   & 1.7   & (1--3) \\
        \noalign{\smallskip}
\EM\,($10^{62}$m$^{-3}$kpc$^{-2}$)& --   &             & 1.9   &            & 0.3  &            & 0.26  &           & --    & \\
$kT_{\rm e}$ (keV)               &  --  &              & 5     & ($>3.7$)   & 3.7  &(2.2--9)    & 5     & ($>3.7$)     & --    & \\
\net ($10^{15}$m$^{-3}$s)        &  --  &              & 0.7   & ($<2.6$)   & 2.5  & (1--5)     & 3.5   & ($< 4.7$) & --    & \\
       \noalign{\smallskip}
N  				 & 0    & (0-0.04)     & 0     & (0.0--0.01)& 0.02 & (0--0.2)   & 0.03  & (0--0.3)  & 0     & (0--0.03)   \\
O                         	 & 0.09 & (0.06--0.12) &0.014&(0.008--0.021)& 0.01 &(0.02--0.1) & 0.012 & (0--0.06) & 0     & (0--0.002)  \\
Ne                        	 & 0.25 & (0.21--0.28) & 0.078&(0.065--0.086)& 0.08&(0.10--0.25)& 0.08  &(0.04--0.5)& 0.04  & (0.03--0.05)\\
Mg                        	 & 0.12 & (0.10--0.15)  & 0.06  &(0.04--0.07)& 0.07 &(0.04--0.14)& 0.13  &(0.03--0.5)& 0.03  & (0.0--0.09) \\
Si                        	 & 0.19 & (0.14--0.24) & 0.15  & (0.10--20) & 0.15 &(0.07--0.24)& 0.16  & (0--0.7)  & 0.3   & (0.1--0.5)\\
S                         	 & 0.3  & (0.0--0.6)   & 0     & (0--0.2)   & 0    &(0--0.4)    & 0     & (0--1.2)  & 0.12  & (0.0--1.0)  \\
Ar                        	 & 8    & (4.0--13)    & 0     & (0--2.5)   & 10   & (5--17)    & 0     & (0--3)    & 0     & (0.0--2.0)  \\
Fe				 & 0.16 & (0.14-0.20)  & 0.12  &(0.09--0.16)& 0.10 &(0.01--0.2) & 0     & (0--$> 10$)&2.8   & (0.1--8)   \\
Fe (hottest component)           &  --  &              & 2     & (1--3)     & 6    & (2--21)    & 2     &  (0--5)   & --    & \\
            \noalign{\smallskip}
\NH ($10^{21}$cm$^{-2}$)  	 & 2.3  & (2.0--2.7)   & 1.6    &(1.3--2.0) & 2.7  & (1.8--4.2) & 2.0   & (0.5--3.5)& 1.3   & (1.0--1.7)\\
$\chi^2_{\nu}$            	 & 4.7  &              & 1.7    &           & 1.2  &            & 1.8   &           & 1.7   & \\
        		\noalign{\smallskip}
        		\hline 
      			\end{tabular} 
		\end{flushleft}
	}
\end{table*}

\begin{figure}[tb]
	%\picplace{12cm}
	\psfig{figure=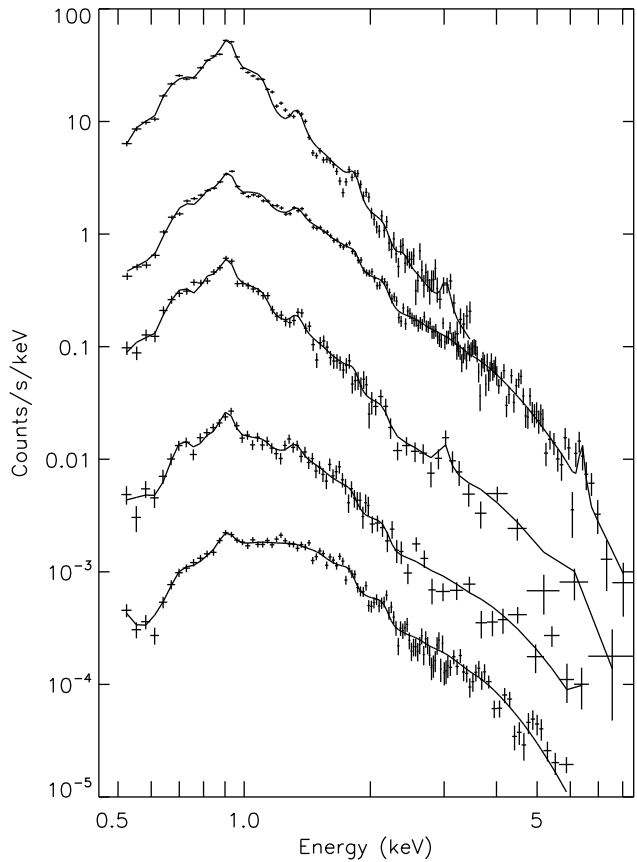,width=8.8cm}
   	\caption[]{
SIS spectra of various regions together with 
NEI model spectra (solid lines). From top to bottom: 
SW Soft ($\times 10$), SW Hard, NE Soft, NE Center ($\times 0.1$), 
NE Hard ($\times 0.005$). \label{spectra}
	}
\end{figure}

\paragraph{Spectral modeling.}
As a starting point for looking for spectral variation over the remnant 
we made hardness ratio maps for all pointings and instruments. Rather than 
showing all ASCA ratio maps, we present here the ROSAT PSPC ratio map
(Fig. \ref{ratio}\footnote{
A color version and other images are available at {\em http://saturn.sron.ruu.nl/$\sim$jaccov/}
}).
Note that with the spatial resolution of ASCA 
these regions were less distinct than in Fig.~\ref{ratio}. 
So the spectral
differences shown in Fig.~\ref{spectra} may in reality be larger.
Because the ratio map does not give a good impression of the morphology of 
the remnant, we also present a PSPC image of RCW\, 86.
Fig.~\ref{image} shows that the X-ray emission extends beyond
the main shell. This was not seen in the HRI image by Pisarski et al. 1984.  
In particular the North{\em west} of the remnant shows faint emission outside 
the main shell. This cannot be
an effect of the point spread function or interstellar scattering, because
then one would expect a similar effect for the North{\em east} of the remnant, 
where the main shell is as bright as in the Northwest.
Furthermore interstellar scattering takes place preferentially at lower 
energies (the photon cross section scales as $E^{-2}$, 
see Predehl \& Schmitt 1995), whereas Fig.~\ref{ratio} shows that the faint
emission in the Northwest is rather hard. 
Finally note that the \HAlpha image (Smith 1997)
and the radio surface brightness profile (Pisarski et al. 1984) show also 
emission outside the main shell. For \HAlpha the emission seems to come
from breakouts from the main shell.

The ratio map shows that we can identify two distinct 
regions in the Southwest (SW). For obvious reasons we will call spectra from 
these regions SW Hard and Soft. The SW Soft region correlates very well with 
the \HAlpha emission (Smith 1997), 
whereas the SW Hard region shows a rough correlation with the radio map
(Kesteven \& Caswell 1987). The SW Soft region is essentially the same as the
bright boomerang-shaped shell in Fig.~\ref{image}, although the SW Hard 
region has some overlap with this shell, suggesting a projection effect.

In the fainter Northeastern (NE) part of the remnant we also find hardness 
variations. The spectra of the hardest region we will tag NE Hard. We 
extracted spectra East of this, including the bright arm (NE Soft)
and more to the center of the remnant (NE Center). As Fig.~\ref{spectra} 
shows the spectra are distinctively different, but all have a characteristic 
peak arising from helium-like Ne around 0.9~keV. The NE spectra are in 
general harder than the SW spectra. The correlation between faintness and
hardness (the softest emission comes from the brightest regions) suggests 
that the shock has decelerated in the densest region compared with the
fainter regions such as the Northeast.

As noted above ASCA has a fairly large point spread function with broad wings
making that the half power diameter is 3\arcmin\ (Tanaka et al., 1994, 
Serlemitsos et al. 1995). 
One might wonder how this affects the spectra in Fig.~\ref{spectra}. 
Most serious is the contamination of the SW Hard spectrum by emission from the
SW Soft region for energies around 1~keV. For higher energies ($>$ 2~keV) it 
is the other way around. As an example, based on the PSPC image and spectra we
estimate that the contribution of the emission from the soft region on the 
spectrum of the hard region is less than 25\% at 1~keV. Such differences are
much smaller than the overall differences between both spectra 
(Fig.~\ref{spectra}).

We fitted the spectra from all regions with one and two temperature component 
NEI models using the spectral code SPEX (Kaastra et al. 1996).
The free parameters were: 
the emission measure (EM) \EM\ (where $d$ is the distance, $n_{\rm H}$ 
and $n_{\rm e}$, the hydrogen and electron number densities and 
$V$ the volume occupied by the plasma);
the ionization parameter \net ($t$ being the time since the plasma 
was heated); the electron temperature \Te;
the interstellar column density \NH\ and the abundances of 
N, O, Ne, Mg, Si, S, Ar and Fe. 
Since in some cases the Fe abundance can be 
estimated from both the Fe-L and Fe-K emission, we left the Fe abundance 
of the hot (i.e. Fe-K emission) and cool component free. In most 
cases the temperature of the hot component was not very well determined; 
in order to prevent temperatures much in excess of what was found by Ginga
(Kaastra et al. 1991) we set an upper limit to the temperature 
of $kT_{\rm e} = 5$~keV.
We have used the interstellar extinction model of Morrison \& McCammon (1983).
Table~\ref{parameters} lists the results.

\begin{figure}[td]
	%\picplace{7cm}
	\psfig{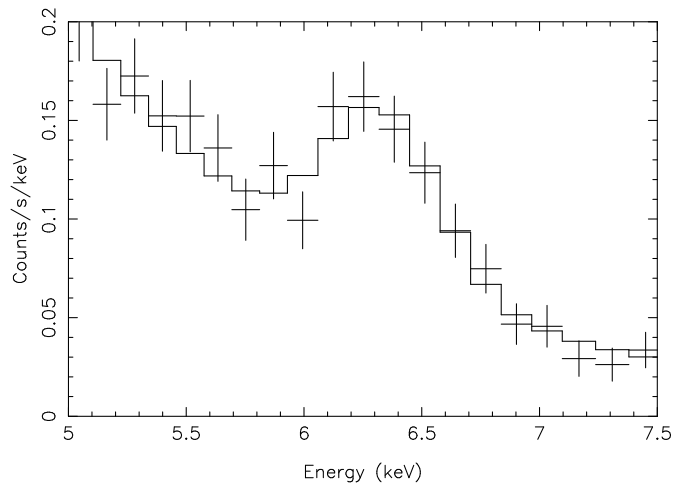}
    	\caption[]{
The Fe-K complex in the South as observed by 
GIS2 and GIS3 (combined in one spectrum). The centroid of the complex
is $6.3\pm0.1$~keV consistent with a fluorescent origin.\label{ironK}
	}
\end{figure}

\paragraph{Abundances.}
There are two remarkable things to 
be noted about the best fitting models. First, the \net\ values are very low, 
in some cases as low as $2\, 10^{15}$m$^{-3}$s, one of the lowest value ever 
reported (\net\ values a factor ten lower were recently reported for some 
regions of the Vela SNR based on ROSAT PSPC spectra, Bocchino et al., 1997). 
Second, the abundances are peculiar. In general, the abundances are
low, except for some cases of Ar and Fe. The high
Fe abundance is not simply an artefact of the model, which is clear from 
Fig.~\ref{ironK}. It shows that the centroid of the Fe-K complex is at 
6.4~keV, indicating a fluorescent origin. The Ar fluorescent emission at 
3~keV is present in both the SIS (Fig.~\ref{spectra}) and the GIS spectra. 
Note that 
fluorescence caused by irradiance is very unlikely in view of the
large column densities and the large radiation field needed.
So are the low abundances real?
What could possibly be the origin of the abundance pattern?
An approach to these problems is 
to analyze the assumptions made in fitting the spectra.

In the spectral code abundances are 
essentially determined by the line to continuum ratios. The continuum is
supposed to be mainly bremsstrahlung from electron-ion collisions,
assuming a Maxwellian electron energy distribution. Furthermore, the
assumption is made that the plasma can be approximated by a 
simple one or two component plasma model assuming constant temperature, 
although in reality temperature and density gradients must be present.
For most SNRs such a model works reasonably well.
A possible extra source of continuum radiation can originate from 
synchrotron radiation arising from shock accelerated electrons with energies
of the order of a few TeV. Such a model has been proposed for SN 1006 
(Koyama et al. 1995). However, in our case this explanation does not work. 
First of all, the SW Soft spectrum is too soft to be dominated by a 
synchrotron component which is usually assumed to have a power law behaviour. 
The second reason is even more compelling: a power law component makes it even 
harder to understand why we see such a relatively strong Fe-K complex around 
6.4~keV.

Another way to deal with underabundances, dust depletion, is
very unlikely in this case because the abundance pattern we find 
is not what one expects it to be in case of dust depletion.
For instance Ne is underabundant, whereas it cannot be depleted in 
grains. Nevertheless, although it does not explain our overall results, 
dust depletion may give an additional effect for some metals (O, Mg, Si, Fe).

One might also think that an increased He/H ratio with respect to 
the solar value results in a stronger continuum emission and thus in a 
smaller line to continuum ratio.
However, this is not the case. To see this, consider two plasmas, 
one with a solar He/H ratio and one for which part of the H has been used to 
make He (mostly $^4$He). 
Because 4 H ions provide more electrons than one He ion, the plasma with the 
solar He/H ratio will emit more bremsstrahlung continuum.

So we probably need a modification of the standard model.
A possibility is that the electron distribution is not Maxwellian.
The continuum emission may partly result from a (non-thermal) tail 
to a relatively  cool ($\simlt 0.05$~keV) electron distribution. 
As a result, metals should be in a lower ionization stage than implied by 
the model fits. Such a model does at least 
qualitatively explain why fluorescent emission indicates different
(i.e. high) abundances.

We have tried to model such a 
distribution by calculating the ionization stages of O for a mix of 
a high ($\sim 5$~keV) and low temperature electron distribution 
($\sim 0.02$~keV). This crude model indicates that 
we can hide an appreciable amount (40\%) of metals as long as \net\ is low 
(e.g. $\simlt 5\, 10^{15}$m$^{-3}$s for a cool to hot ratio of 4 to 1). 
Note that an appreciable amount of cool
electrons is not unlikely in the light of recent UV observations of SN\, 1006
(Laming et al. 1996), which indicated  $T_{\rm e}/T_{\rm ion}\simlt 0.05$.
As an indication: a shock velocity for RCW\, 86
of 700~km/s (Long \& Blair 1990) without electron--ion equilibration gives 
$kT_e = 0.0005$~keV; Laming et al.'s result suggests 
$kT_e \simlt 0.05$~keV.
Clearly the effects of a non-Maxwellian electron distribution need
further modeling. It is furthermore unclear why all the spectra of RCW 86
share so many characteristics, whereas they are taken from very different
regions.

Up to now we did not address the issue how projection or contamination 
effects may be responsible for the unusual ASCA RCW\, 86 spectra. 
In our opinion it is
very hard to explain the near absence of line emission solely by projection 
effects without invoking an unconventional interpretation. 
In a model that uses projection effects the observer should see a 
superposition of a spectrum with a ``normal'' line
to continuum ratio and a spectrum with an almost complete absence of lines.
So the difficulty of how to suppress line emission remains unaltered in such a 
model. In fact we cannot exclude that the ASCA SIS spectra are superpositions 
of two opposite cases, but we want to point out that it is remarkable that 
such extremely smooth spectra exist.

\begin{figure*}[td]
	%\picplace{7cm}
	\psfig{figure=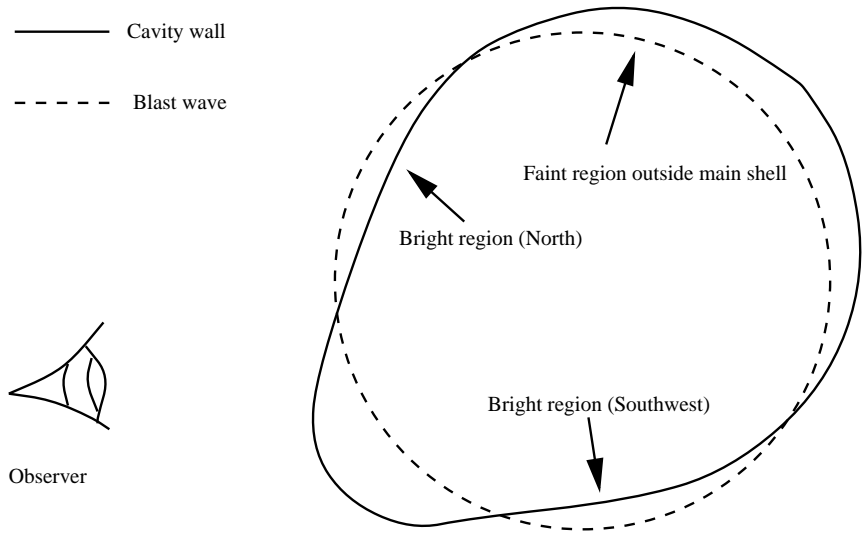,height=7cm,angle=-90}
    	\caption[]{
A sketch of our interpretation of the structure of RCW 86 as a cavity 
explosion. We have drawn the blastwave here as a sphere, but in reality the
shape of the blastwave will be affected when it hits the cavity wall. Bright 
are those regions where the blastwave has hit the cavity wall.
\label{sketch}
	}
\end{figure*}

\paragraph{Evidence for a cavity explosion.}
Pisarski et al. 1984 and Claas et al. 1989 reported that there is a large 
density contrast between the North and South. In agreement with their results
we find a (post shock) electron density of 
$(1-10)\, 10^6 d^{-0.5}_{\rm kpc}$\densm\ 
for the SW Soft region and $0.2\, 10^6 d^{-0.5}_{\rm kpc}$~\densm\ for the 
NE region ($d_{\rm kpc}$ is the distance in kpc). 
The range in densities for the SW Soft region comes from the uncertainty in 
the emitting volume, which can be anything within a factor of 10 of
$10^{50}d^3_{\rm kpc}$m$^3$. The estimate for the NE region was based on 
a filling factor of 0.25, a coverage by the Northeastern SIS pointing of 0.2 
and a total EM of $5\, 10^{62}$m$^{-3}$kpc$^{-2}$.
The optical emission tells an even more interesting story: \HAlpha observations
of non-radiative shocks in the North give a pre-shock density of 
$0.2\, 10^6$~\densm 
(Long \& Blair 1990), whereas spectra taken from the Southwest indicate
densities of more than 10$^8$~\densm\ (Leibowitz \& Danziger 1983).

The density contrast may have an accidental cause, 
but for a number of reasons we think that RCW\,86 
makes a strong case for being the result of a cavity explosion. 
In particular 
we think that the cavity wall has not been reached by the shock wave all 
over the remnant. 
With a cavity we mean here a tenuous region in the ISM blown by
one or more stellar winds (Weaver et al., 1977), such a cavity is 
surrounded by a denser region, the cavity wall, consisting of 
swept-up material.
The reasons for our hypothesis are:
(1) the low density in the North, 
given the fact that RCW 86 is situated near the galactic plain; 
(2) the faint emission outside the main shell
in Fig. ~\ref{image}, most notably in the Northwest, implying that the wall 
of the cavity has not been reached in all directions (see Fig.~\ref{sketch}); 
(3)  Fig. 3 in Claas et al. (1989) shows that the radial X-ray 
profile of the Northeastern part of the remnant has too much emission coming 
from the shell of the SNR as compared to the Sedov model;
(4) the more or less egg-shape 
reminds one of a cavity made by a star which was moving from North to 
South with respect to the interstellar medium (Weaver et al. 1977, 
see their Fig. 7), or alternatively the cavity may have been blown by several
stars; 
(5) the gap in the Northern shell recalls the breakouts seen in 
numerical simulations of cavity SNRs by Franco et al. (1991).
This is even more striking in the \HAlpha picture by Smith (1997) where
ahead of the gap there is an enhancement in the \HAlpha emission;
(6) a recent encounter of the blastwave with the cavity wall makes it more 
likely that complete electron heating has not yet occured and may also explain
why the \net\ values are rather low: most of the matter has only recently been 
shocked when the blastwave hit the cavity wall.

In order to clarify point 2 we made a sketch of the
structure of RCW 86 as we envision it (Fig.~\ref{sketch}). 
Note that the figure gives a side view and that we do not know the 
inclination, so it might also be the Southwestern region that is closer to us.
It is clear from our sketch that we do not know the exact shape of 
the cavity, because we only see those parts of the cavity wall that are lit up
by the blastwave. However, it is also clear 
that if the cavity has an egg-like shape, the remnant will also have an 
egg-like shape, but possibly less elongated. Fig.~\ref{sketch} also clarifies
the relativily low emission from the center of RCW 86 (point 3): 
along the line of sight the cavity wall has not yet been reached by the 
blastwave.

\section{Conclusions}
We have presented an analysis of ASCA spectra of RCW\,86, which we modeled by
NEI models. In general these models fit the data rather well, but indicate 
an underabundance of metals. On the other 
hand, we need over-abundances to explain the fluorescent emission of Ar and Fe
in some parts of the remnant.
The fitted models indicate a very low \net\ value.

We suggest that the peculiarities of the spectra may be the result of
a non-Maxwellian electron distribution. If the bulk of the electrons 
has a very low temperature, most of the ions should be in a low ionization
stage (e.g. the Li-like stage), whereas
a supra-thermal tail may dominate the X-ray continuum and explain
the Ar-K and Fe-K fluorescent emission. Note that our density calculations 
underestimate the true densities if the electron distribution is non-thermal.

Other SNRs may exhibit similar effects, although perhaps to a lesser
extent. 
In that respect it is 
interesting that Miyata et al. (1994) and Miyata (1996) report similar 
underabundances for the Cygnus Loop.

We suggest that RCW\,86 is the result of a cavity explosion. The
morphology of the remnant is a good indication for that and even
suggests that the cavity may have been blown by more than one star. 
This is interesting 
in view of the new distance estimate by Rosado et al. (1996), placing RCW\,86 
at the distance of an OB association at 2.5~kpc (Westerlund 1969). 
If the association is real, RCW\,86 is probably the result of a SN type II
explosion.

\begin{acknowledgements}
It is a pleasure to thank Dr. Y. Tanaka for fruitful discussions and his 
useful suggestions. We also thank the referee Dr. R. Strom for his critical 
reading of the article.
This research has made use of data obtained through the High Energy
Astrophysics Science Archive Research Center Online Service, provided
by the NASA/Goddard Space Flight Center. This work was financially
supported by NWO, the Netherlands Organization for Scientific
Research.
\end{acknowledgements}

\end{document}